\documentclass[journal,10pt]{IEEEtran}

\usepackage[pdftex]{graphicx}
\usepackage{amsmath}
\usepackage{amsfonts}
\usepackage{amssymb}
\usepackage{subfigure}
\usepackage{color}
\usepackage{float}
\usepackage{color}
\usepackage{cuted}

\begin{document}
\title{Band-pass Magnetic Tunnel Junction based Magnetoresistive Random Access Memory}
\author{Abhishek Sharma, Ashwin Tulapurkar and Bhaskaran Muralidharan
	\thanks{Abhishek Sharma, Ashwin Tulapurkar and Bhaskaran Muralidharan are with the Department
		of Electrical Engineering, IIT Bombay, Powai, Mumbai-400076, India. e-mail: (bm@ee.iitb.ac.in)}
}
\maketitle
\begin{abstract}
We propose spin transfer torque--magnetoresistive random access memory (STT-MRAM) based on magneto-resistance and spin transfer torque physics of band-pass spin filtering. Utilizing the electronic analogs of optical phenomena such as anti-reflection coating and resonance for spintronic devices, we present the design of an STT-MRAM device with improved features when compared with a traditional trilayer device. The device consists of a superlattice heterostructure terminated with the anti-reflective regions sandwiched between the fixed and free ferromagnetic layers.  Employing the Green’s function spin transport formalism coupled self-consistently with the stochastic Landau-Lifshitz-Gilbert-Slonczewski equation, we present the design of an STT-MRAM based on the band-pass filtering having an ultra-high TMR ($\approx3.5\times10^4\%$) and large spin current. We demonstrate that the STT-MRAM design having band-pass spin filtering are nearly 1100\% more energy efficient than traditional trilayer magnetic tunnel junction (MTJ) based STT-MRAM. We also present detailed probabilistic switching and energy analysis for a trilayer MTJ and band-pass filtering based STT-MRAM. Our predictions serve as a template to consider the heterostructures for next-generation spintronic device applications.     
\end{abstract}
\begin{IEEEkeywords}
Magnetic tunnel junction, spin-transfer torque, STT-MRAM, Resonant tunneling.
\end{IEEEkeywords}
\IEEEpeerreviewmaketitle
\section{Introduction}
\begin{figure}[h!]
	\centering
	\subfigure[]{\includegraphics[width=2.3in]{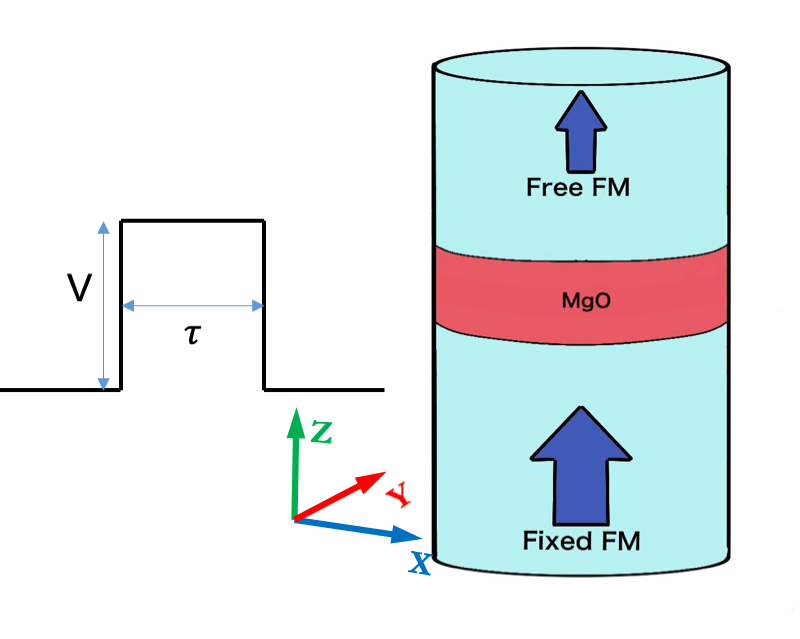}
 	}\subfigure[]{\includegraphics[width=1in]{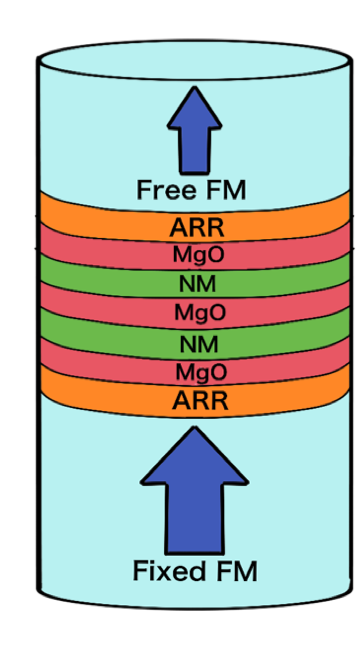}}		
	\subfigure[]{\includegraphics[width=1.3in]{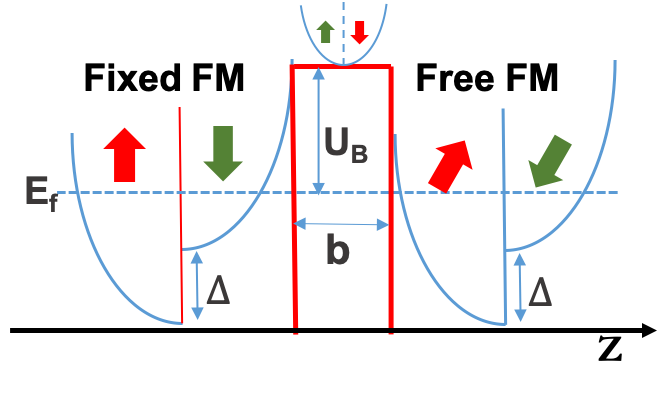}	
	}\subfigure[]{\includegraphics[width=2.2in]{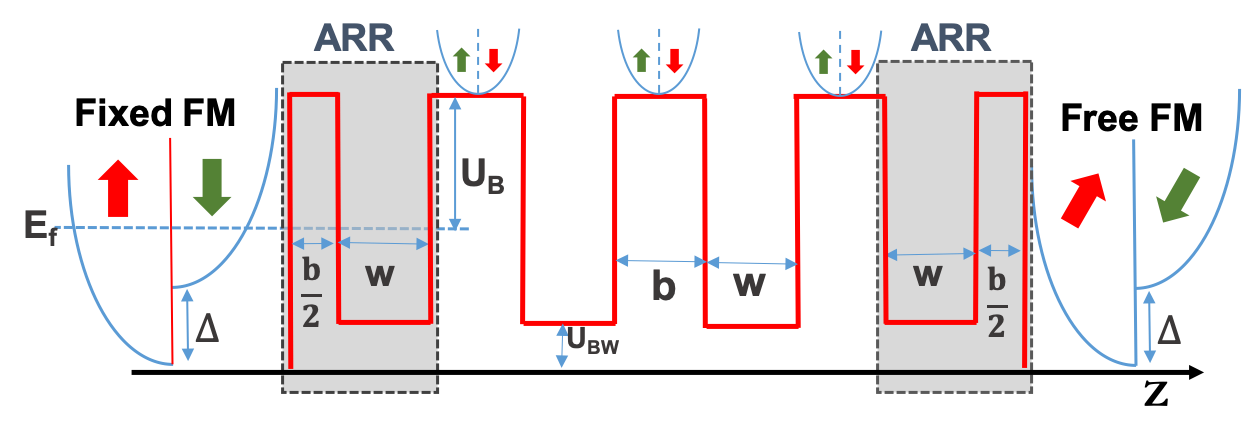}}
	\caption{Device schematics: (a) A trilayer magnetic tunnel junction (MTJ) based MRAM device having an MgO barrier separating fixed and free FM layers, (b) a  BPMTJ based MRAM device with 3-barriers or 2-quantum wells having alternating layers of MgO (red) barrier and normal metal (green) well along with anti-refection regions sandwiched between the free and fixed FM layers. A voltage pulse with varying bias and pulse width has been applied to evaluate switching characteristics. (c) Band diagram of a trilayer MTJ and (d) the BPMTJ device along the direction of magnetization of the fixed layer FM.}
	\label{chap6_device_design}
\end{figure}

magnetoresistive random access memory (MRAM) is a promising candidate of next-generation RAM due to its comparable performance with dynamic random access memory (DRAM) and integrability on the existing CMOS technology\cite{Yu2016,Endoh2016}. The MRAM  has a distinct advantage of the non-volatility and inexhaustible write endurance over commercially used DRAM. The non-volatile nature of the MRAM has the potential to club the storage and random access memory part of the existing computer architecture. A magnetic tunnel junction (MTJ) is one of the suitable candidates for the MRAM device due to the tunnel magnetoresistance (TMR) and spin transfer torque (STT) effects. An MTJ device consists of an insulator barrier sandwiched between two ferromagnetic(FM) contacts. The magnetization of the bottom FM layer is fixed called as the fixed FM layer, whereas the magnetization of upper FM layer is free to rotate under the influence of magnetic field or STT is known as the free FM layer. The resistance of an MTJ device depends on the relative orientation of the magnetization of the free and fixed FM layer due to spin-dependent tunneling\cite{Butler2001} and is quantified by TMR ratio: 
\begin{equation}
TMR=\frac{R_{AP}-R_{P}}{R_{P}}\times100
\end{equation}
where, $R_P$ and $R_{AP}$ are the resistance in the parallel configuration (PC) and anti-parallel configuration (APC) of magnetization of FMs. In an MTJ based MRAM device, information is stored in the relative orientation of the fixed and free ferromagnets (FM), as shown in the Fig.~\ref{chap6_device_design}(a). The PC and APC of magnetization of FMs can represent `$0$' and `$1$', respectively. The TMR effect of an MTJ device is utilized to read the magnetization state of the free FM layer relative to the fixed FM.\\
\indent The magnetization direction of the free layer can be changed by applying a static magnetic field higher than the coercivity of the free FM.  Alternatively, the magnetization of a nano-magnet can be switched using a spin-polarized current through spin transfer torque  (STT) effect\cite{Slonczewski1996,Berger1996}. The spin of an electron is a quantized angular momentum which can be transferred to the local magnetization of the free FM. When a spin-polarized current incident on the FM, the traverse component of the spin current gets absorbed in the FM which exerts a torque on the magnetization of the FM known as STT\cite{Ralph1,brat} effect. The STT is a non-conservative quantity which can act either as a damping term by increasing the magnetic relaxation of the system or as an anti-damping term depending on the direction of the spin current. When STT acts as an anti-damping term, it compensates for the magnetic relaxation process inherent in the system making the magnetization of the FM unstable and eventually flipping the direction of magnetization.\\
\indent The information on an MRAM cell can be written either by utilizing the STT effect or via magnetic field lines emerging out of the current carrying line in an MRAM chip. 
But magnetic field switching based architecture increases the size and power consumption of the MRAM chip\cite{Apalkov2016}. STT-MRAM architecture is more compact and energy efficient. The time required to write an MRAM cell (i.e., writing latency)  and energy consumed to write information are the major performance indicators for an MRAM device. The writing latency, in turn also increases the energy consumption during the switching process. While a typical MTJ based STT-MRAM has the advantage of being non-volatile at the same time comparable performance to the commercially available DRAM but energy consumption during the writing process in a typical STT-MRAM is nearly an order of magnitude higher than DRAM\cite{Yu2016}.\\
\indent There have been many efforts in the research community to reduces the energy consumption and writing latency of MRAM via reducing the cross-sectional area of the MTJ device\cite{Gajek2012,Gottwald2015,Perrissin2018}, as well as utilizing the perpendicular surface-anisotropy between MgO and COFeB free FM layer\cite{Tudu2017} to reduce the current switching requirements. This has led to the development of perpendicular(p)-MTJ (Fig.~\ref{chap6_device_design}(a)) which have a very smaller switching bias due to the absence of the demagnetization field. These efforts have been primarily focused on the free layer FM engineering, but alternatively one can envision device structural changes to improve the STT-MRAM performance. When it comes to device structural engineering penta-layered MTJs, have been explored for different applications such as magnetic sensors\cite{Sharma2016} and spin torque oscillators\cite{Sharma2017,Sharma_aip_2018}. But these penta-layered MTJs are not suitable for STT-MRAM applications due to the asymmetric spin current profile.\\
\indent In this work, we try to alleviate the above described challenges faced by an MTJ based STT-MRAM via  device structural engineering. We propose an energy efficient STT-MRAM device design based on the band-pass spin filtering physics. In our earlier work\cite{Sharma2018}, we have shown various heterostructure designs to realize the band-pass spin filtering. This work is geared toward the practical device realization an energy efficient STT-MRAM based on band-pass spin filtering.
\begin{figure}[!h]
	\centering
	\includegraphics[width=3.4in]{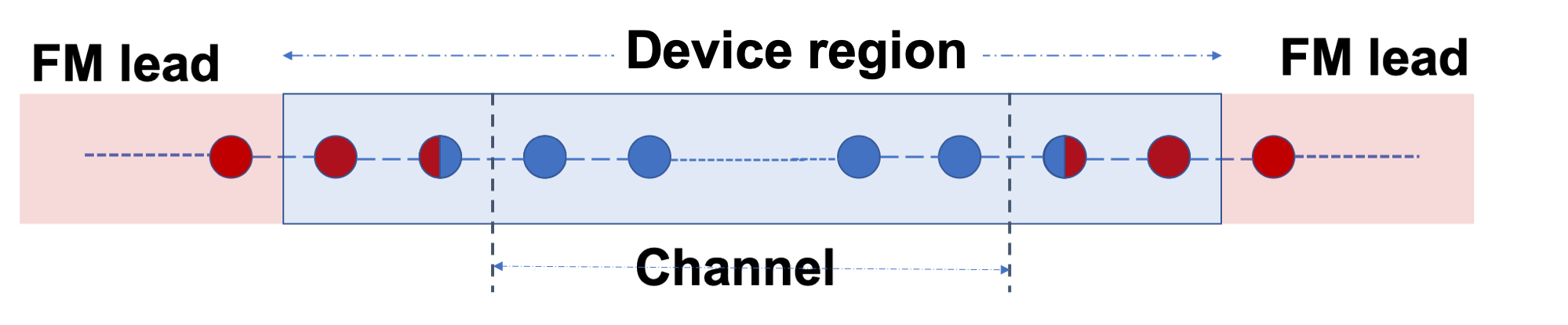}		
	\caption{Tight binding chain of atoms for each transverse mode, the device region includes at least one atom from both the FMs. Red dot represents the FM lattice points, blue dot represents the channel lattice points and interface atom have equal contribution from channel and adjacent FM. }
	\label{atom_chain}
\end{figure}
\section{Simulation details and device design}
\indent We show in the Fig.~\ref{chap6_device_design}(a) and (b) p-MTJ and p-BPMTJ based MRAM cell. The magnetization of the free and fixed FM is perpendicular to the plane of device.  We have employed the non-equilibrium Green's function (NEGF) \cite{datta1} spin transport formalism coupled with the stochastic Landau-Lifshitz-Gilbert-Slonczewski (LLGS)\cite{Slonczewski1996,Berger1996} equation to describe the magnetization dynamics of the free FM. \\
\indent We show in the Fig.~\ref{chap6_device_design}(c) and (d) band diagram of a trilayer MTJ and the BPMTJ structure. The trilayer MTJ has a layer of MgO between the magnets while the BPMTJ has a superlattice of MgO-normal metal (NM)-MgO-NM-MgO sandwiched between the anti reflective region (ARR) terminated with fixed and the free magnets. The ARR region (see section \ref{ARC_details}) consists of same NM with same width that of superlattice quantum well NM and a MgO barrier with half of the width that of superlattice barrier as shown in the Fig.~\ref{chap6_device_design}(d).\\
\indent We describe the details of  the spin NEGF transport formalism which has been used to capture the device physics. We start with the device Hamiltonian matrix $[H]$ and the  energy resolved spin dependent single particle Green's function matrix $[G(E)]$ is given by:
\begin{equation}
[G(E)] = [EI-H-\Sigma]^{-1}
\label{green_fun}
\end{equation}
\begin{eqnarray}
[\Sigma]=[\Sigma_T]+[\Sigma_B]
\label{Sigma}
\end{eqnarray}
where, the device Hamiltonian matrix,  $[H]=[H_0]+[U]$, comprises of the device spin dependent tight-binding matrix, $[H_0]$ and the applied potential profile matrix ,$[U]$, in real space, $[I]$ is the identity matrix with the dimensionality of the device Hamiltonian. The quantities $[\Sigma_T]$ and $[\Sigma_B]$ represent the spin dependent self-energy matrices\cite{datta2} of the bottom and top FM contacts evaluated within the tight-binding framework\cite{yanik,deepanjan}.\\
\indent We have used spin dependent single band effective mass Hamiltonian to evaluate the Green's function of the device. The device region consists of the channel region plus at least one point from both the FM contacts as shown in the Fig.~\ref{atom_chain}. Both the FM contacts have been described using the Stoner model of FM\cite{Ralph1} with FM exchange energy ($\Delta$), effective mass ($m^*_{FM}$) and Fermi energy ($E_f$). The spin dependent Hamiltonian of the device in the tight binding matrix for each transverse mode can be written as:
\begin{strip}
	\begin{equation}
	[H_0] = \left ( \begin{array}{cccccccc} 
	\alpha_{B}^{FM} & \beta_B^{FM} & 0 & . &.&.&.&.\\ 
	\beta_B^{\dagger FM} & \alpha_I^{B\rightarrow C} & \beta_C & 0&.&.&.&.\\
	0 & \beta_C^{\dagger} & \alpha_C & \beta_C &0&.&.&. \\ 
	. &0 & \beta_C^{\dagger}  &  \alpha_C &\beta_C&0&.&.\\
	. &.&0& . & .&.&0&.\\
	. &.&.& 0 & .&.&.&0\\
	.& .&.&.&0&\beta_C^{\dagger} & \alpha_I^{C\rightarrow T} & \beta_T^{FM} \\
	& . & . &.&.&0& \beta_T^{\dagger FM}&\alpha_{T}^{FM}  \\ 
	\end{array} \right ),
	\label{eq:TB_H0}
	\end{equation}
\end{strip}
where, $\alpha_{B}^{FM}$  spin dependent on-site energy matrix of the bottom FM is given by:
\begin{equation*}
\alpha_{B}^{FM}= \left ( \begin{array}{cc}
2t_{FM}+E_t& 0\\
0& 2t_{FM}+E_t+\Delta
\end{array} \right )
\end{equation*} 
$\beta_B^{FM}$, spin dependent coupling matrix of the bottom FM is given by:
\begin{equation*}
\beta_B^{FM}= \left ( \begin{array}{cc}
-t_{FM}& 0\\
0& -t_{FM}
\end{array} \right )
\end{equation*} 
$\alpha_I^{B\rightarrow C}$,  spin dependent on-site energy matrix of the interface between bottom FM and channel region is given by:
\begin{equation*}
\alpha_I^{B\rightarrow C}= \left ( \begin{array}{cc}
\tau_I& 0\\
0& \tau_I +\Delta
\end{array} \right )
\end{equation*} 
where, $\tau_I=t_{FM}+t_{C}+0.5(E_t+E_t\left(m_{FM}^*/m_C^*)\right)$ is onsite energy term for interface atom.\\
$\beta_C$, spin dependent coupling matrix of the channel region is given by:
\begin{equation*}
\beta_C= \left ( \begin{array}{cc}
-t_{C}& 0\\
0& -t_{C}
\end{array} \right )
\end{equation*} 
$\alpha_{C}$,  spin dependent on-site energy matrix of the channel region is given by:
\begin{equation*}
\alpha_C= \left ( \begin{array}{cc}
2t_{C}& 0\\
0& 2t_{C}
\end{array} \right )
\end{equation*} 
$\alpha_I^{C\rightarrow T}$,  spin dependent on-site energy matrix of the interface between the channel and top FM is given by:
\begin{equation*}
\alpha_I^{C\rightarrow T}= \left ( \begin{array}{cc}
\tau_I& 0\\
0& \tau_I+\Delta
\end{array} \right )
\end{equation*} 
$\beta_T^{FM}$, spin dependent coupling matrix of the top FM is given by:
\begin{equation*}
\beta_T^{FM}= \left ( \begin{array}{cc}
-t_{FM}& 0\\
0& -t_{FM}
\end{array} \right )
\end{equation*} 
$\alpha_{T}^{FM}$,  spin dependent on-site energy matrix of the top FM is:
\begin{equation*}
\alpha_{T}^{FM}= \left ( \begin{array}{cc}
2t_{FM}+E_t& 0\\
0& 2t_{FM}+E_t+\Delta
\end{array} \right )
\end{equation*} 
where, $t_{FM}=\hbar^2/2m^*_{FM}a^2$ and $t_{C}=\hbar^2/2m^*_{C}a^2$ are coupling parameters between each site in the FM and channel regime, respectively. $E_t$ is the transverse mode energy, $m^*$ is the effective mass and $a$ represents the lattice spacing. 
In general, the magnetization of the free/top layer FM is not always along the $\hat{z}$ and a unitary transformation is applied to the $\alpha_{T}^{FM}$ and $\alpha_I^{C\rightarrow T}$\cite{yanik}.This unitary transformation matrix is given by:
\begin{equation*}
u_{trans}=\begin{pmatrix}
cos~\theta/2 & sin~\theta/2~e^{-i\phi}\\
-sin~\theta/2~e^{i\phi} & cos~\theta/2\\
\end{pmatrix}
\end{equation*} 
where, the $\theta$ is the relative angle between the magnetization of the free and fixed FM along the $z$-direction given by $\theta=cos^{-1}(m_z)$. Similarly,  $cos(\phi)=m_x/sin(\theta)$ and $sin(\phi)=m_y/sin(\theta)$  where $m_x$, $m_y$ and $m_z$ is the $x$, $y$ and $z$-component of the unit vector along the magnetization of the free FM layer. If there are $N$ number of lattice points in the device region as shown in the Fig. \ref{atom_chain}, then all the matrices would be of size $2N\times2N$ due to spin dependency.\\ 
$\left[ \Sigma_B\right]$ and $\left[ \Sigma_T\right]$ of Eq.~\ref{green_fun} represent self-energy matrices of the bottom (fixed) and top (free) FM layers respectively \cite{datta2}. $\left[ \Sigma_B\right]$ is nonzero only for the first 2$\times$2 block which is given by:
\begin{equation*}
\sigma_1=\begin{pmatrix}
-t_{FM}~e^{ik^{\uparrow}_B a} & 0\\
0 & -t_{FM}~e^{ik^{\downarrow}_B a}
\end{pmatrix}
\end{equation*}
and $\left[ \Sigma_T\right]$ is nonzero only for the last 2$\times$2 block which is given by:
\begin{equation*}
\sigma_N=u_{trans}\begin{pmatrix}
-t_{FM}~e^{ik^{\uparrow}_B a} & 0\\
0 & -t_{FM}~e^{ik^{\downarrow}_B a}
\end{pmatrix}u^{\dagger}_{trans}
\end{equation*}
where, $k^{\uparrow,\downarrow}_{B,T}$ are related to the spin dependent $E-k$ relation inside the FM as shown in the Fig.~\ref{chap6_device_design}(c) and (d) by the following equations:
\begin{align*}
E^{\uparrow} &= E_t+2t_{FM}\left(1-cos~k^{\uparrow}_{B,T} a\right)\\
E^{\downarrow} &= E_t+\Delta +2t_{FM}\left(1-cos~k^{\downarrow}_{B,T} a\right).\\
\end{align*}
\indent A typical matrix representation of any quantity $[A]$ defined above entails the use of the matrix element $A(z,k_x,k_y,E)$, indexed on the real space $z$ and the transverse mode space $k_x,k_y$. We follow the uncoupled transverse mode approach to account for the finite cross-section,with each transverse mode indexed as $k_x,k_y$ evaluated by solving the sub-band eigenvalue problem \cite{deepanjan,lunds,salah2} with transverse mode energy $E_t=\frac{\hbar^2k_x^2}{2m_{FM}}+\frac{\hbar^2k_y^2}{2m_{FM}}$. \\
\indent The electron density in NEGF formalism is diagonal elements of the energy resolved electron correlation matrix  $[G^n(E)]$ which is given by:
\begin{equation}
[G^n]= \int dE [G(E)] [\Sigma^{in}(E)] [G(E)]^{\dagger}	\label{G_n}
\end{equation}
\begin{equation}
[\Sigma^{in}(E)]=[\Gamma_T(E)]f_T(E)+[\Gamma_B(E)]f_B(E) \label{Sigma_in}
\end{equation}
Here, $[\Gamma_T(E)]=i\left ([\Sigma_T(E)]- [\Sigma_T(E)]^{\dagger} \right )$ and $[\Gamma_B(E)]=i\left( [\Sigma_B(E)]-[\Sigma_B(E)]^{\dagger} \right )$ are the spin dependent broadening matrices \cite{datta2} of the top and bottom contacts. The Fermi-Dirac distributions of the top/free FM and bottom/fixed FM contacts are given by $f_T(E)$ and $f_B(E)$ respectively. The charging matrix, $[U]$, is obtained via assuming a linear drop of the potential in the barrier region and no drop inside the metal region and subjected to the boundary conditions, $U_{FixedFM}=-qV/2$ and $U_{FreeFM}=qV/2$, with $V$ being the applied voltage. \\
\indent The matrix element of the current operator $\hat{I}_{op}$ representing the current between two lattice points\cite{datta1} i.e $N-1$ and $N$ (at the interface of channel and free FM) is given by:
\begin{equation}
{I}_{op,N-1,N}=\frac{i}{\hbar}\left(H_{N-1,N}G^{n}_{N,N-1}-(H_{N-1,N}G^{n}_{N,N-1})^{\dagger}\right)
\end{equation}
the current operator $\hat{I}_{op}$ is a 2$\times$2 matrix in spin space. The charge current $I$ is given by 
\begin{equation}
I =q \int dE \text{ Real [Trace(}\hat{I}_{op}\text{)]}
\end{equation}
and spin current $I_S$ is given by
\begin{equation}
I_{S\sigma} =q \int dE \text{ Real [Trace(}\sigma_S\cdot\hat{I}_{op}\text{)]}
\end{equation}
where, $\sigma_s$ is Pauli spin matrices along x, y, z directions.
We use the Landau-Lifshitz-Gilbert-Slonczewski (LLGS) equation to calculate the magnetization dynamics of the free layer in the presence of an applied magnetic field and spin current\cite{Slonczewski1996,brat}:
\begin{equation}
\begin{split}
\left( 1+\alpha^{2}\right) \frac{\partial \hat{m}}{\partial t} = -\gamma \hat{m} \times (\vec{H}_{eff}+\vec{h_{fl}}) \\- \gamma \alpha \left( \hat{m} \times ( \hat{m} \times (\vec{H}_{eff}+\vec{h_{fl}}))\right)  \\
- \frac{\gamma\hbar}{2qM_SV} \left((\hat{m}\times(\hat{m}\times\vec{I_S}))-\alpha(\hat{m}\times\vec{I_S})\right) \label{eq:llgs}
\end{split}	
\end{equation}
where $\hat{m}$ is the unit vector along the direction of magnetization of the free magnet, $\gamma$ is the gyromagnetic ratio of the electron, $\alpha$ is the Gilbert damping parameter, $\vec{H}_{eff} = \vec{H}_{app} + H_{k \perp} m_{z}\hat{z}$ is the effective magnetic field with $\vec{H}_{app}$ being the applied external field, $H_{k \perp}$ being the anisotropy field.\\
\indent The spin current can be resolved as:
\begin{equation}
\vec{I_S}=I_{S,m}\hat{m}+I_{S,\parallel}\hat{M}+I_{S,\perp}\hat{M}\times\hat{m}  \label{eq:spin_current},
\end{equation}
where $\hat{M}$ is the unit vector along the direction of magnetization of the fixed magnet, using the equation \eqref{eq:spin_current}, equation \eqref{eq:llgs} can be re-casted ( as $\alpha$ is small, ignoring the $\alpha(\hat{m}\times\vec{I_S})$ term) as :
\begin{equation}
\begin{split}
\left( 1+\alpha^{2}\right) \frac{\partial \hat{m}}{\partial t} = -\gamma \hat{m} \times (\vec{H}_{eff}+\vec{h_{fl}})\\ - \gamma \alpha \left( \hat{m} \times ( \hat{m} \times (\vec{H}_{eff}+\vec{h_{fl}}))\right)  \\
- \frac{\gamma\hbar}{2qM_SV} \left((\hat{m}\times(\hat{m}\times I_{S,\parallel}\hat{M}))+ (\hat{m} \times I_{S,\perp} \hat{M})  \right) \label{eq:llgs_split}
\end{split}	
\end{equation}
It can be inferred from the equation \eqref{eq:llgs_split} that the spin current  $I_{S\parallel}$ along  $\hat{M}$ acts as a damping/anti-damping term depending upon direction of the current and is known as Slonczewski spin transfer torque term and the $I_{S\perp}$ along $\hat{M}\times\hat{m}$ act as a magnetic field is known as field like term.\\ 
\indent We have also taken into account the thermal noise in the form of magnetic field fluctuations $\vec{h_r}$ in the LLGS equation with the following statistical properties \cite{Garcia-Palacios1998}
\begin{equation}
\langle h_{fl,i}(t)\rangle = 0, \langle h_{fl,i}h_{fl,j}(s)\rangle = 2D\delta_{i j}\delta(t-s)
\end{equation}
where i and j are Cartesian indices, and $\langle \rangle$ represents the ensemble average. The strength of the fluctuation $D$ is given by  
\begin{equation}
D=\frac{\alpha}{1+\alpha^2}\frac{k_BT}{\gamma\mu_0M_SV}
\end{equation}\\ 
where, $\mu_0$ is the free space permeability constant, $k_B$ is the Boltzmann constant, T is the temperature of the magnetic layer.\\   
\indent The critical current require for STT switching in the macro-spin assumption is given by 
\begin{equation}
Ic=\frac{2e\alpha}{\hbar} M_sV \left(H_k+\frac{H_d}{2} \right),
\end{equation}
where $M_S$ is the saturation magnetization, $V$ is the volume of the free FM layer, $\alpha$ is the damping constant, $H_k$ is in-plane anisotropy and $H_d$ is the demagnetization field of the free FM. The critical current for switching is proportional to demagnetization field ($H_d$) which is usually an order higher in value than $H_k$. The value of demagnetization field of the free layer can be reduced or changed by introduction of interfacial perpendicular magnetic anisotropy (PMA)\cite{Tudu2017}. The PMA facilitates the magnetization in the free FM to align in the perpendicular direction to the film-plane. The effective perpendicular anisotropy in the presence of demagnetization field and interface PMA is given by:
\begin{equation}
H_{k \perp}=\frac{2\sigma}{M_st} -H_d
\end{equation}
where $\sigma$ is the interface PMA factor and $t$ is the thickness of the free FM layer. The thickness of the free FM should be smaller than a critical value\cite{Ikeda2010} such that interracial-PMA is larger than demagnetization field. The MTJ devices with perpendicular magnetization of the free and fixed FMs are classified as p-MTJ and have lower switching biases. The critical thickness of CoFeB can also be increased by suitable design engineering of the free FM layer\cite{Worledge2011}. In our simulation, we have used CoFeB as the FM with the free layer thickness 1.3nm\cite{Ikeda2010} with perpendicular uni-axial anisotropy $H_{k \perp}=3.3k Oe$\cite{Gajek2012}. In our simulation, we have used the typical parameters of the magnetization dynamics as $\alpha$ = 0.01, the saturation magnetization, $M_S=1150$ emu/cc, $\gamma$ = 17.6 MHz/Oe. The cross-sectional area of all the devices considered is $0.25$$\pi$ $\times$ $30^2$ nm\textsuperscript{2} such that magnetization dynamics of the free FM can be captured via macro-spin model\cite{Sato2014}. The thermal stability factor also known as ferromagnetic energy barrier ($\Delta_{Eb}=\frac{H_kM_sV}{2k_bT}$) between the two stable states for the free FM is $\Delta_{Eb}\approx42k_bT$.  \\
\indent We use the CoFeB parametric tight binding Hamiltonian with exchange splitting $\Delta = 2.15$ eV and Fermi energy $E_f = 2.25$eV. The effective mass of MgO barrier is $m_{OX} = 0.18m_e$, the NM quantum well is $m_{NM} = 0.9m_e$ and that of FM contact is $m_{FM} = 0.8m_e$\cite{deepanjan}, where $m_e$ is the free electron mass. The barrier height of the CoFeB-MgO interface is $U_B = 0.76$ eV above the Fermi energy \cite{kubota,deepanjan}. We have taken conduction band offset between the NM and FM band edge $U_{BW} = 0.5$ eV. In our BPMTJ device design barrier width of $1.2$nm is chosen such that half of the barrier width is $0.6$nm which is the minimum amount of MgO that can be deposited reliably\cite{Deac2008}. We have used the NM with a width of $3.5 \AA$ which lies well within the current fabrication limits\cite{Ryu2013,Yang2015}.\\
\section{Characteristics of p-MTJ and p-BPMTJ}
\begin{figure}[tb!]
	\centering
	\subfigure[]{\includegraphics[width=1.75in]{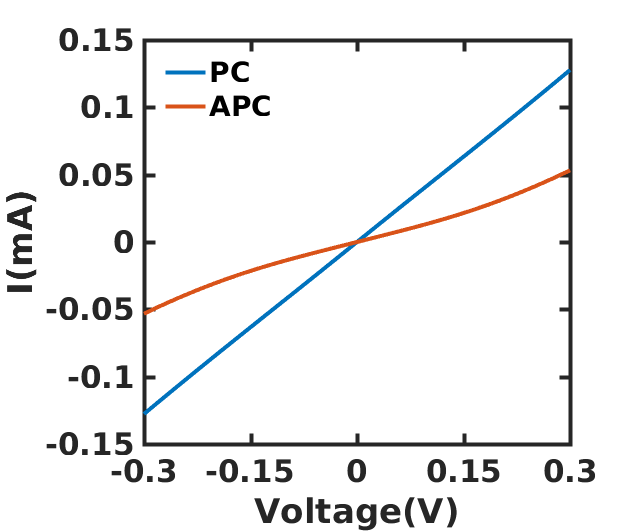}
	}\subfigure[]{\includegraphics[width=1.8in]{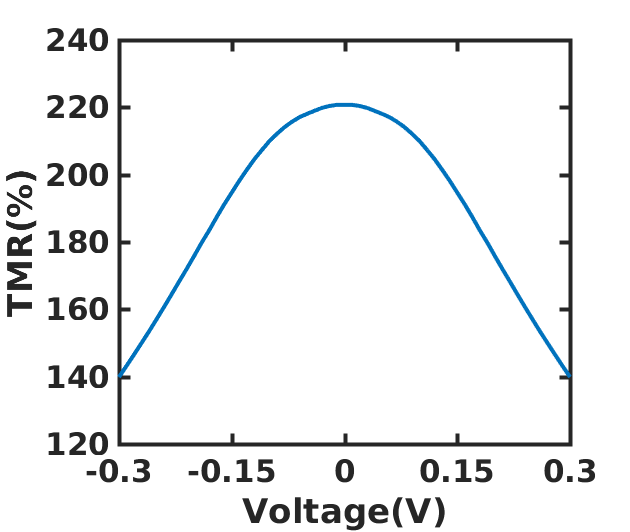}}		
	\subfigure[]{\includegraphics[width=1.8in]{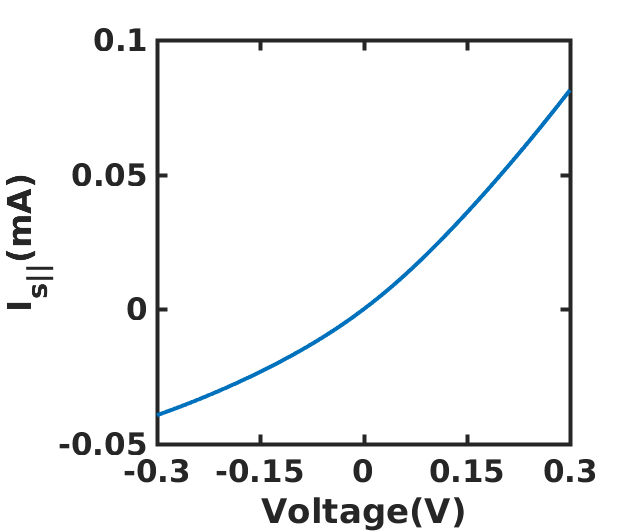}	
	}\subfigure[]{\includegraphics[width=1.8in]{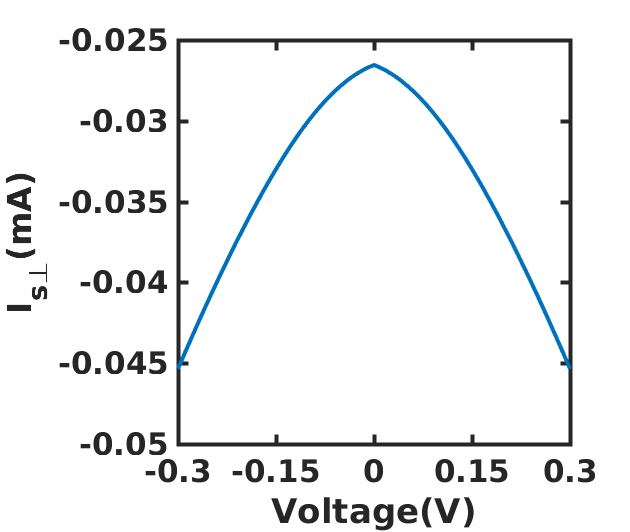}}
	\caption{p--MTJ device characteristics: (a) I-V characteristics in the PC and the APC, (b)  TMR variation with bias voltage, (c) variation of $I_{S\parallel}$ (Slonczewski term) and (d)  variation of $I_{S\perp}$ (field-like term) with applied voltage in the perpendicular configuration of the free and fixed FMs.}
	\label{MTJ_MRAM_IV}
\end{figure}
\indent Figure~\ref{MTJ_MRAM_IV}(a) shows the current-voltage (I-V) characteristics of a trilayer p-MTJ device in the PC and APC. The TMR variation with the voltage for a trilayer device is shown in the Fig.~\ref{MTJ_MRAM_IV}(b). We show in Fig.~\ref{MTJ_MRAM_IV}(c), the variation of the Slonczewski term ($I_{S\parallel}$) of the spin current with bias voltage. Figure~\ref{MTJ_MRAM_IV}(d) shows the variation of the field-like term\cite{butler} ($I_{S\perp}$) of the spin current with voltage bias. The field-like term has a non-vanishing part at zero-bias known as dissipationless spin current. It represents the exchange coupling between the fixed and free FM due to the tunnel barrier \cite{Slonczewski1989}.\\
\begin{figure}[tb!]
	\centering
	\subfigure[]{\includegraphics[width=1.75in]{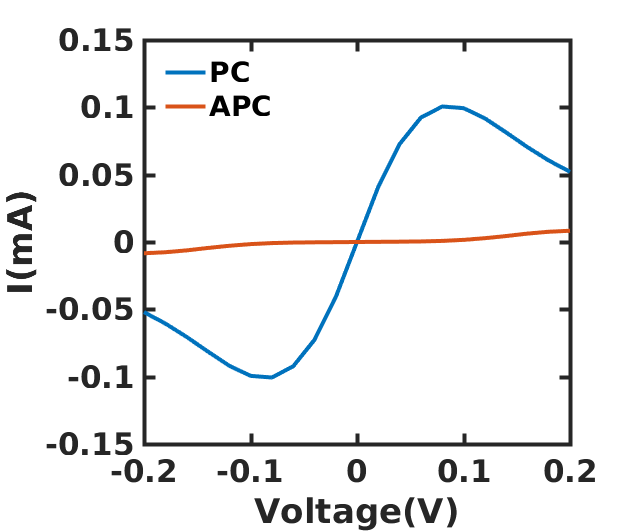}		
	}\subfigure[]{\includegraphics[width=1.8in]{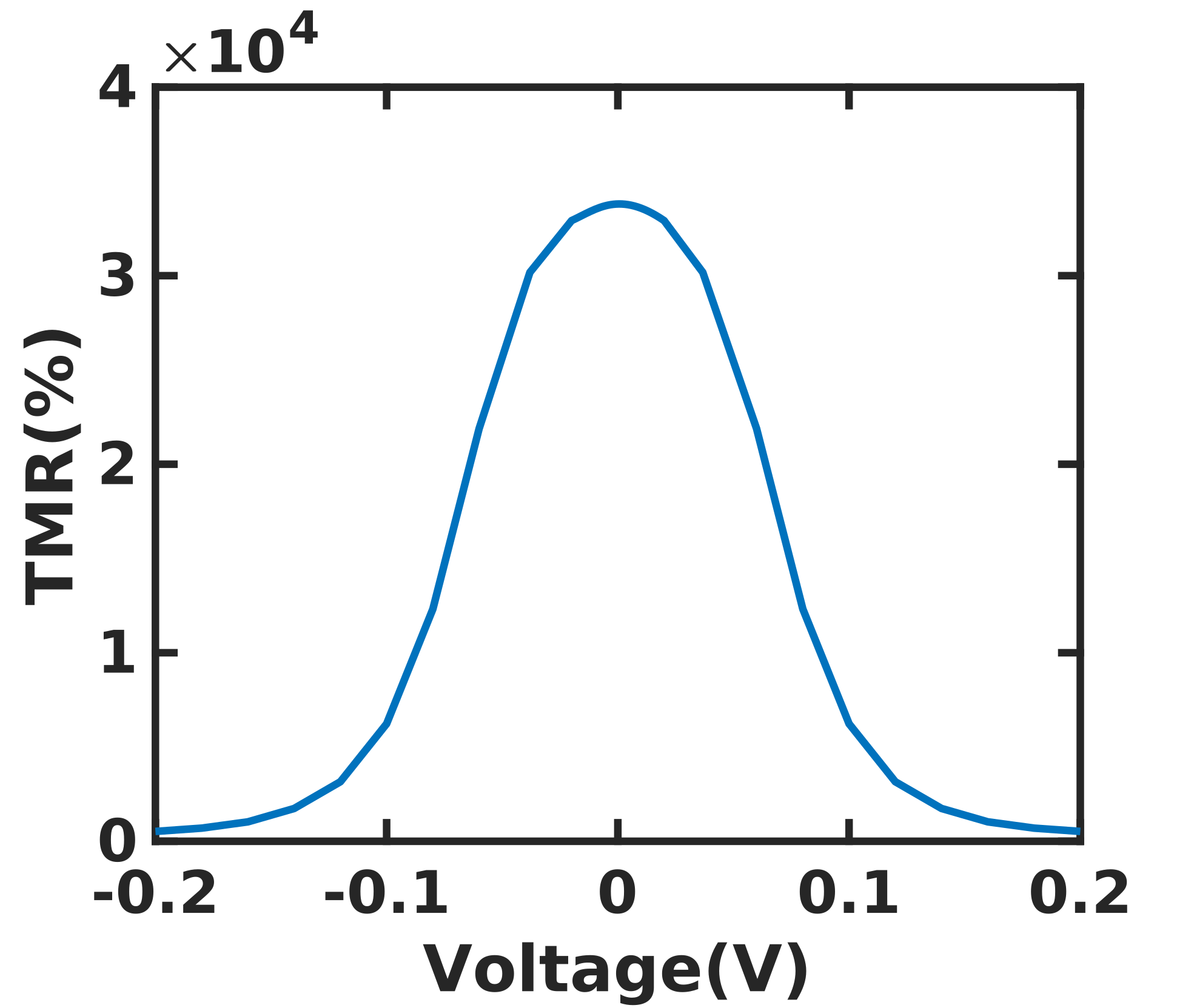}}
	\subfigure[]{\includegraphics[width=1.75in]{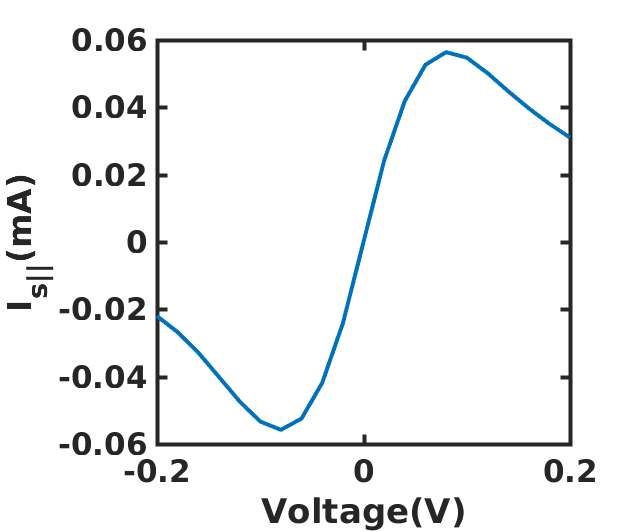}	
	}\subfigure[]{\includegraphics[width=1.8in]{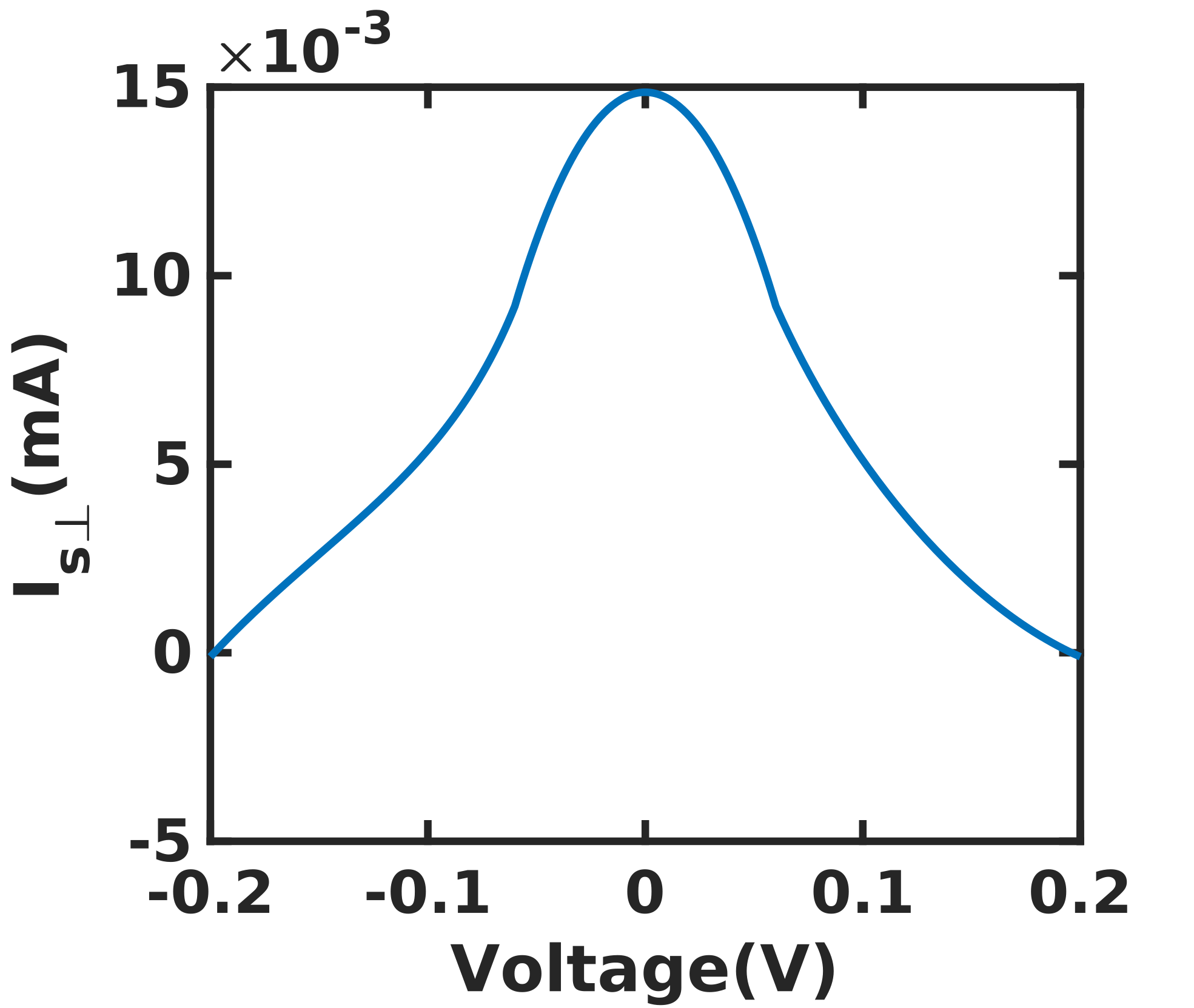}}
	\caption{BPMTJ device characteristics: (a) I-V characteristics in the PC and the APC, (b)  TMR variation with bias voltage, (c)  variation in $I_{S\parallel}$ (Slonczewski term) and (d)  variation in $I_{S\perp}$ (field-like term) with applied voltage in the perpendicular configuration of the free and fixed FMs.}
	\label{ARC_SLTMR_MRAM_V}
\end{figure}
\indent We show the I-V characteristics of p-BPMTJ with a 3-barrier/2-quantum-well structure in Fig.~\ref{ARC_SLTMR_MRAM_V}(a) in the PC and APC. The resonant conduction in the PC and off-resonant conduction in APC results in an ultra-high TMR as shown in the Fig.~\ref{ARC_SLTMR_MRAM_V}(b). We show in the Fig.~\ref{ARC_SLTMR_MRAM_V}(c), the variation of $I_{S\parallel}$ (Slonczewski term) of the spin current with the voltage bias. The nearly symmetric behavior of the Slonczewski term in the p-BPMTJ around zero bias may enable a near symmetric switching voltage in p-BPMTJ based MRAM. We show in the Fig.~\ref{ARC_SLTMR_MRAM_V}(d), the $I_{S\perp}$ (field like term) variation with the voltage bias. The physics of selective band-pass spin filtering in the p-BPMTJ structure provides a large spin current in comparison to the trilayer MTJ. This act as a motivation to analyses the performance of p-BPMTJ based MRAM and compare it with existing p-MTJ based MRAM. \\
\section{Physics of anti-reflection and spin filtering} \label{ARC_details}
\begin{figure}[tb!]
	\centering
	\subfigure[]{\includegraphics[width=1.75in]{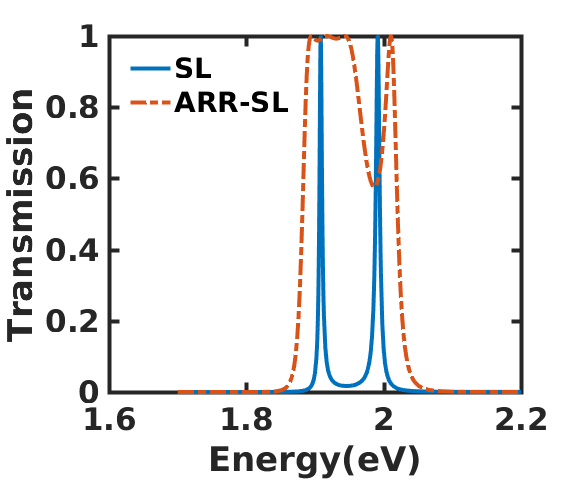}}		
	\subfigure[]{\includegraphics[width=1.75in]{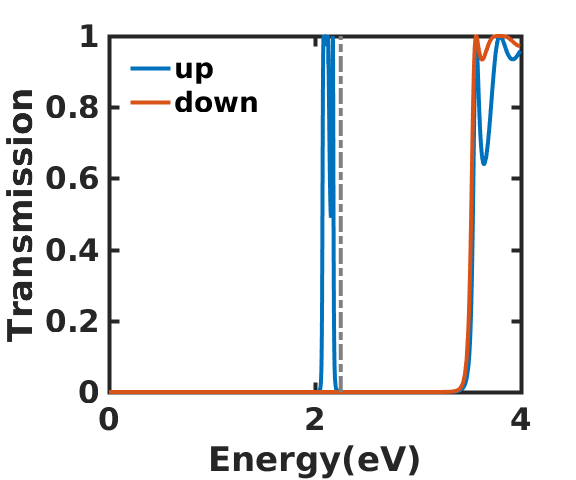}	
	}\subfigure[]{\includegraphics[width=1.8in]{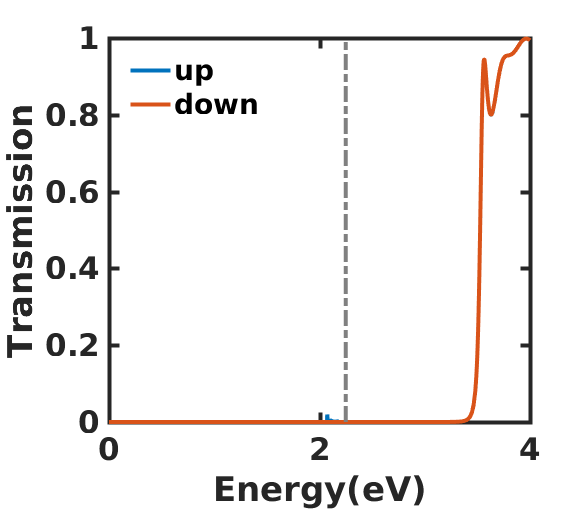}}
	\caption{(a) Transmission spectra of SL and SL with anti-reflective region. Spin resolved transmission spectra for BPMTJ device at transverse mode energy $E_t=0.1$eV for (b) the PC (c) APC at V=0.}
	\label{ARC_phy}
\end{figure}

In this section, we describe the physics of band-pass spin filtering, which leads to ultra-high TMR and large spin current in the BPMTJ device. The band-pass spin filtering is realized by 2-quantum well superlattice sandwiched between anti-reflection regions (ARR) (Fig.~\ref{chap6_device_design}). The 2-quantum well superlattice structure gives resonant peaks in the transmission spectra as shown in the Fig.~\ref{ARC_phy}(a)(Blue). The ARR consist of a quantum well similar to that of SL quantum well and same barrier with width half of the that of SL barrier\cite{Pacher2001,Sharma2018} (Fig.~\ref{chap6_device_design}(b)\&(d)). It acts as an electronic analog of anti-reflection coating resulting in a band of transmission as shown in Fig.~\ref{ARC_phy}(a)(Red).\\
\indent We plot the spin resolved transmission spectra for transverse mode $E_t=0.1$eV in Fig.~\ref{ARC_phy}(b) \& (c) in the PC and APC of the fixed and free FM. It can be inferred from the Fig.~\ref{ARC_phy}(b) that in the PC there is up-spin selective transmission channel for electronic conduction near Fermi level. Whereas in the APC both the up-spin and down-spin channels are blocked near Fermi energy due to band alignment of FMs. This results in a large difference in current conduction of the PC and APC configuration (Fig.~\ref{ARC_SLTMR_MRAM_V}(a)) leading to an ultra-high TMR and large spin current in the BPMTJ structure.
\section{Switching probabilities and switching energy of MRAM device}
\begin{figure}[tb!]
	\centering
	\subfigure[]{\includegraphics[width=1.8in]{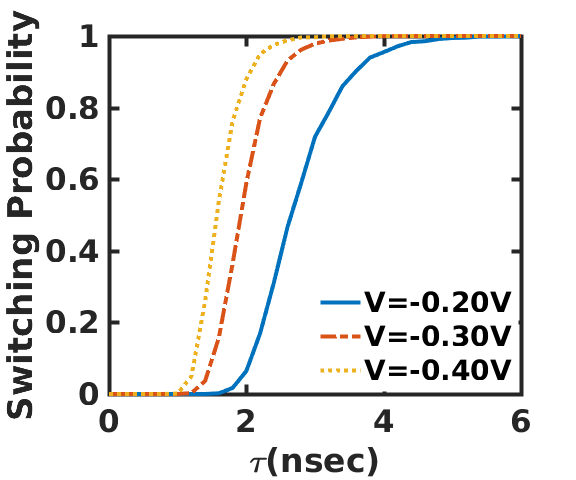}		
	}\subfigure[]{\includegraphics[width=1.8in]{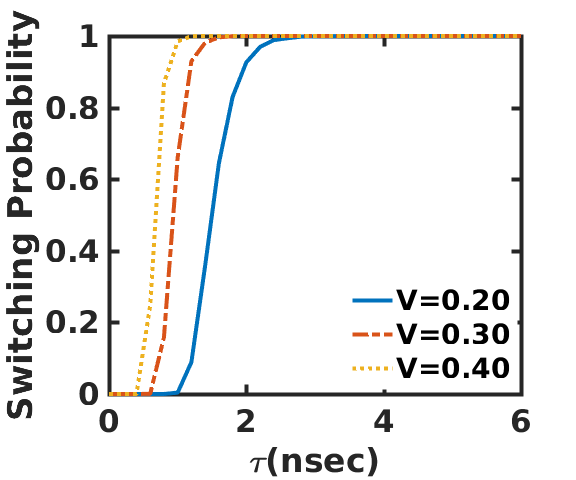}}
	\caption{Switching probability of the p-MTJ based MRAM cell for (a) the PC to APC and (b) the APC to PC switching of free FM}
	\label{MTJ_SW_P_tau}
\end{figure}

\begin{figure}[tb!]
	\centering
	\subfigure[]{\includegraphics[width=1.8in]{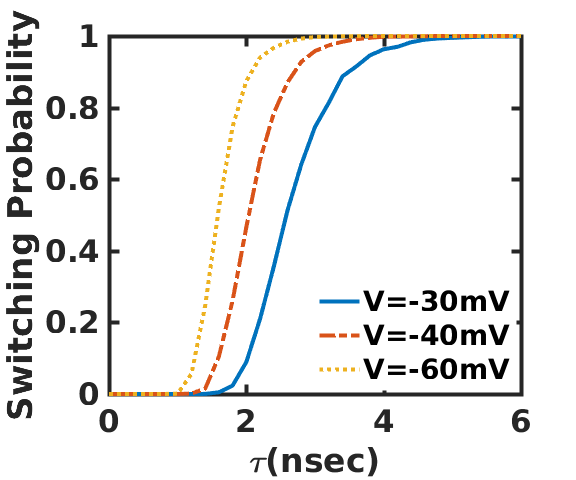}		
	}\subfigure[]{\includegraphics[width=1.8in]{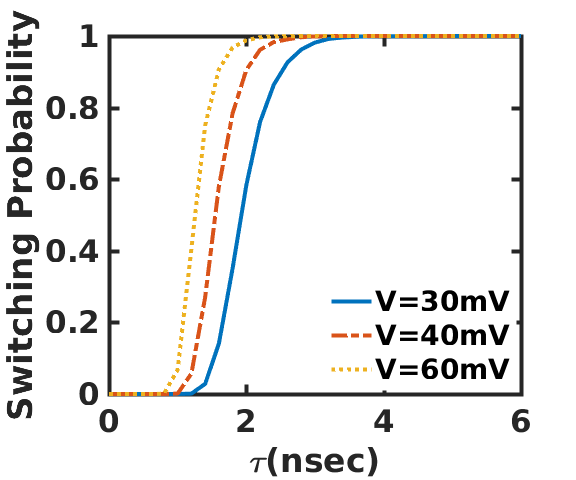}}
	\caption{Switching probabilities of p-BPMTJ based MRAM cell from (a) the PC to APC and (b) the APC to PC switching of the free FM}
	\label{ARFPMTJ_SW_P_tau}
\end{figure}

\begin{figure}[tb!]
	\centering
	\subfigure[]{\includegraphics[width=1.84in]{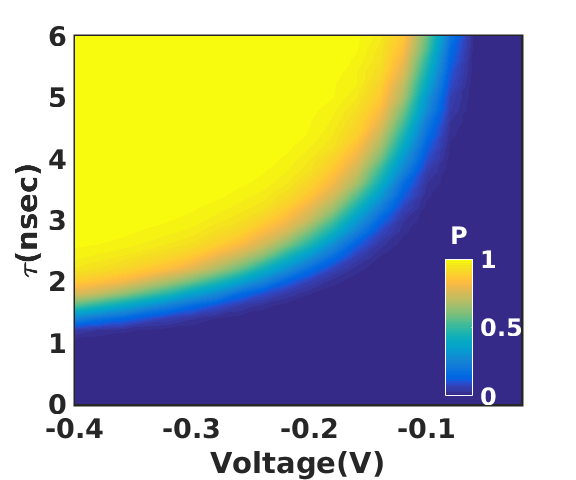}		
	}\subfigure[]{\includegraphics[width=1.74in]{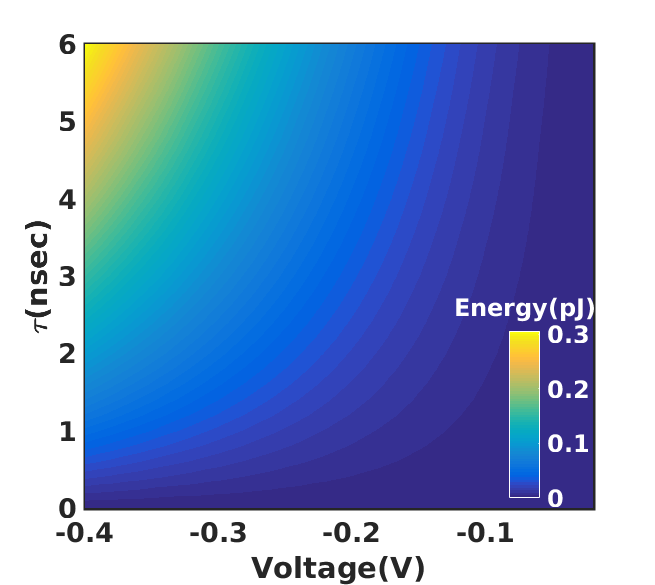}}
	\subfigure[]{\includegraphics[width=1.84in]{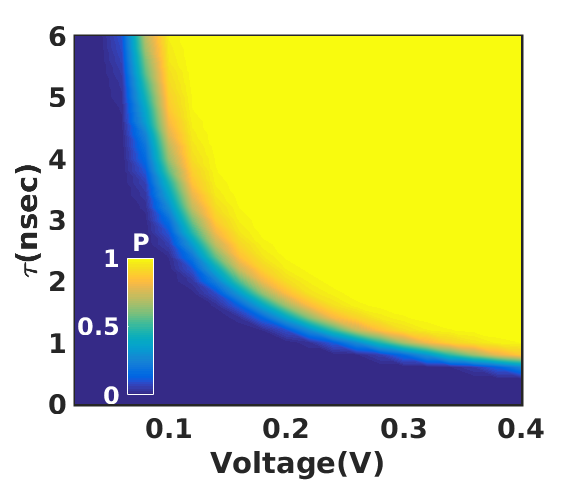}	
	}\subfigure[]{\includegraphics[width=1.73in]{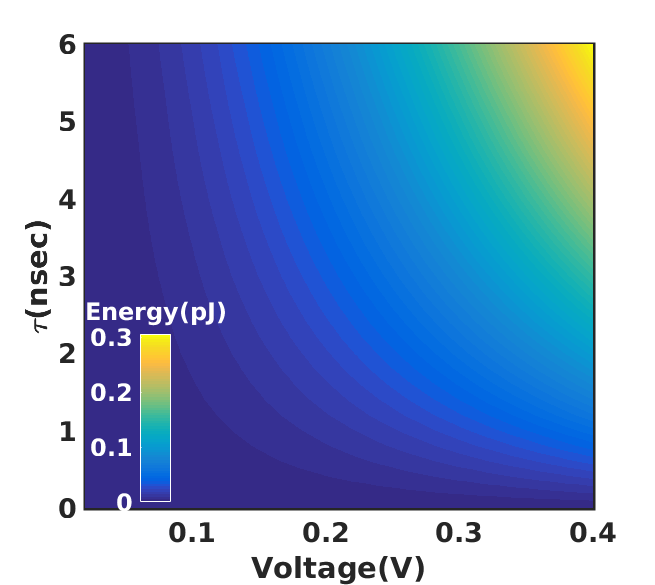}}
	\caption{Voltage-pulse width diagram of the p-MTJ based MRAM for (a) the switching probabilities and (b) the switching energy form the PC to the APC, (c) the switching probabilities and (d) the switching energy form the APC to the PC of the free and fixed FMs.}
	\label{MTJ_SW_P_E_VV_tau}
\end{figure}

\begin{figure}[tb!]
	\centering
	\subfigure[]{\includegraphics[width=1.75in]{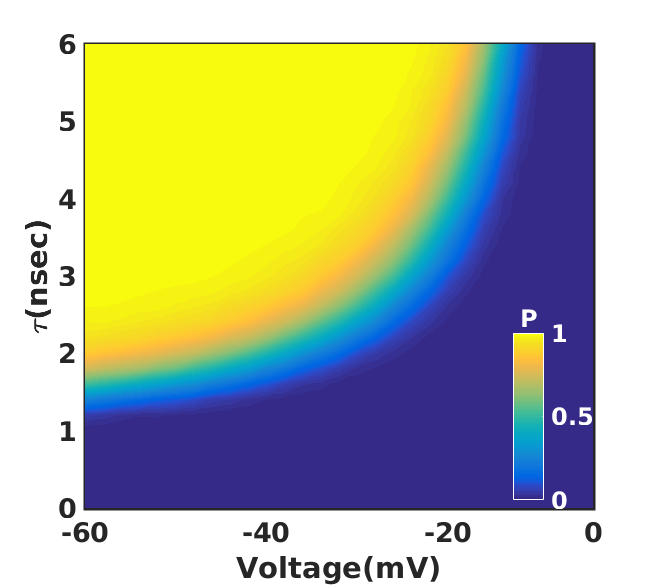}		
	}\subfigure[]{\includegraphics[width=1.75in]{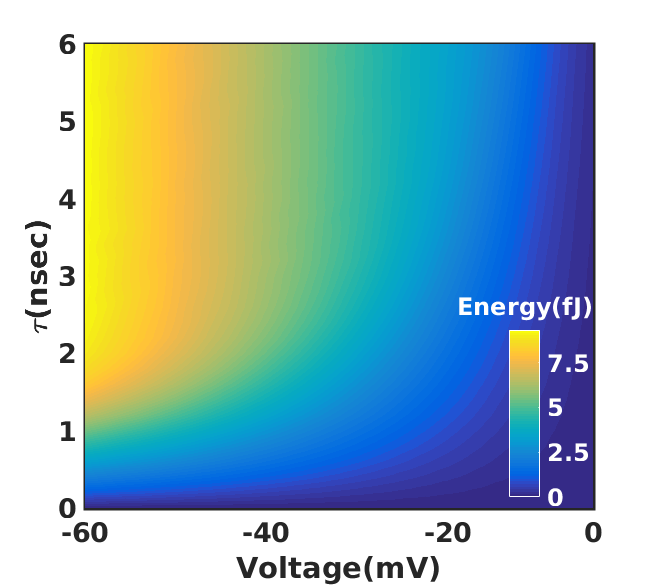}}
	\subfigure[]{\includegraphics[width=1.75in]{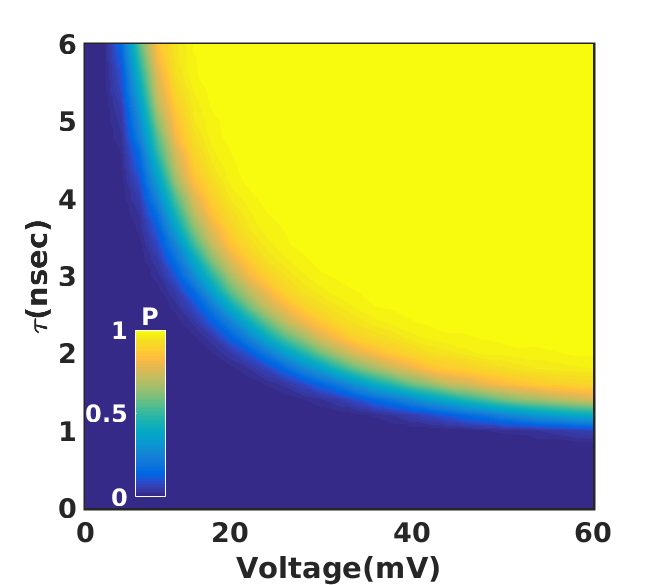}	
	}\subfigure[]{\includegraphics[width=1.75in]{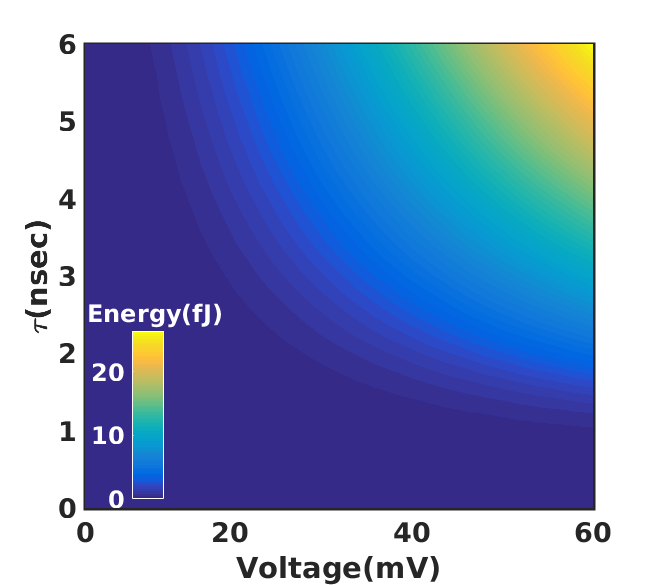}}
	\caption{Voltage-pulse width diagram of the p-BPMTJ based MRAM for (a) the switching probabilities and (b) the switching energy form the PC to the APC, (c) the switching probabilities and (d) the switching energy form the APC to the PC of the free and fixed FMs.}
	\label{ARC_SW_P_E_VV_tau}
\end{figure}

\begin{figure}[tb!]
	\centering
	\subfigure[]{\includegraphics[width=1.75in]{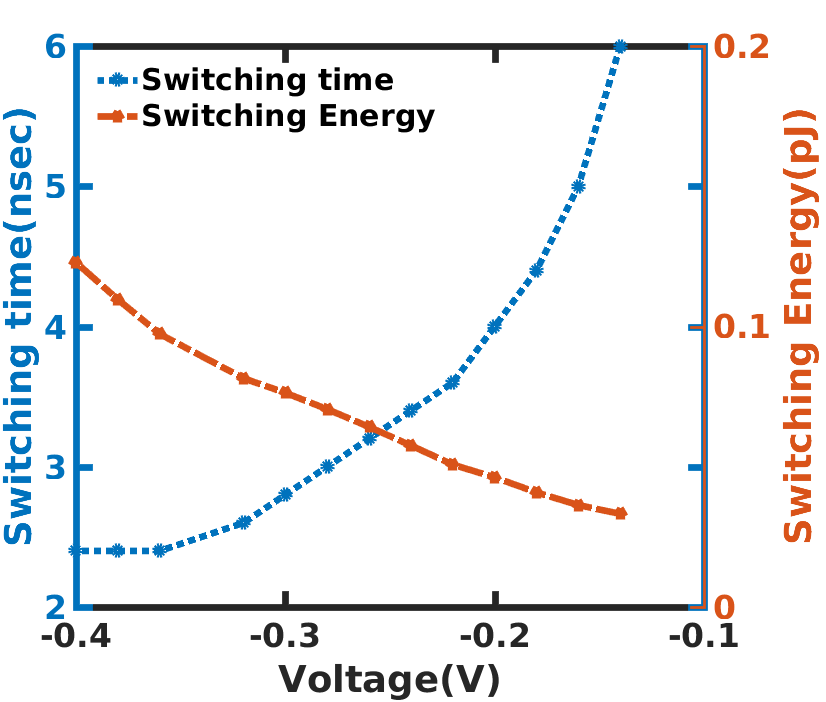}		
	}\subfigure[]{\includegraphics[width=1.75in]{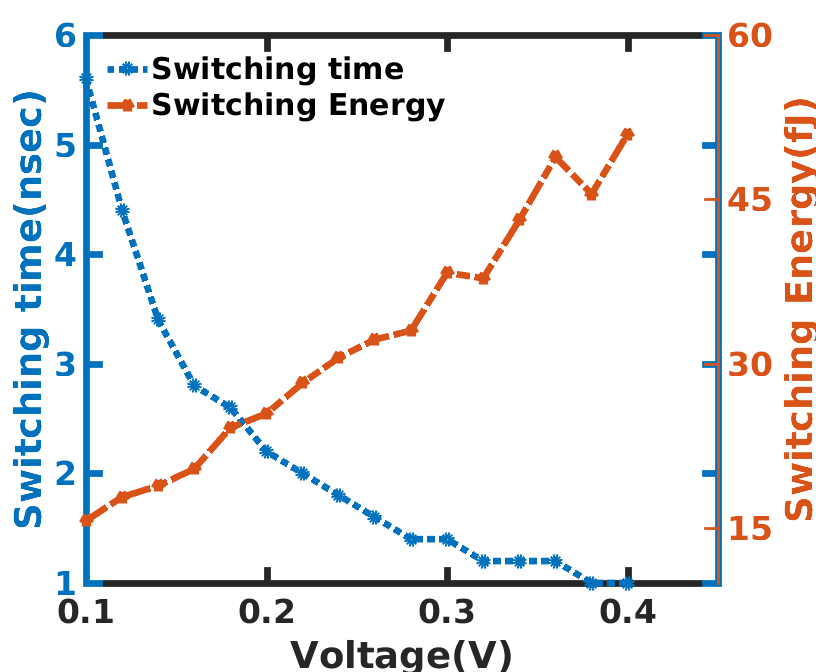}}
	\caption{Switching time and switching energy of p-MTJ based MRAM cell from (a) the PC to APC and (b) the APC to PC of the fixed and free FMs.}
	\label{MTJ_SW_P_E_VV}
\end{figure}

\begin{figure}[bt!]
	\centering
	\subfigure[]{\includegraphics[width=1.75in]{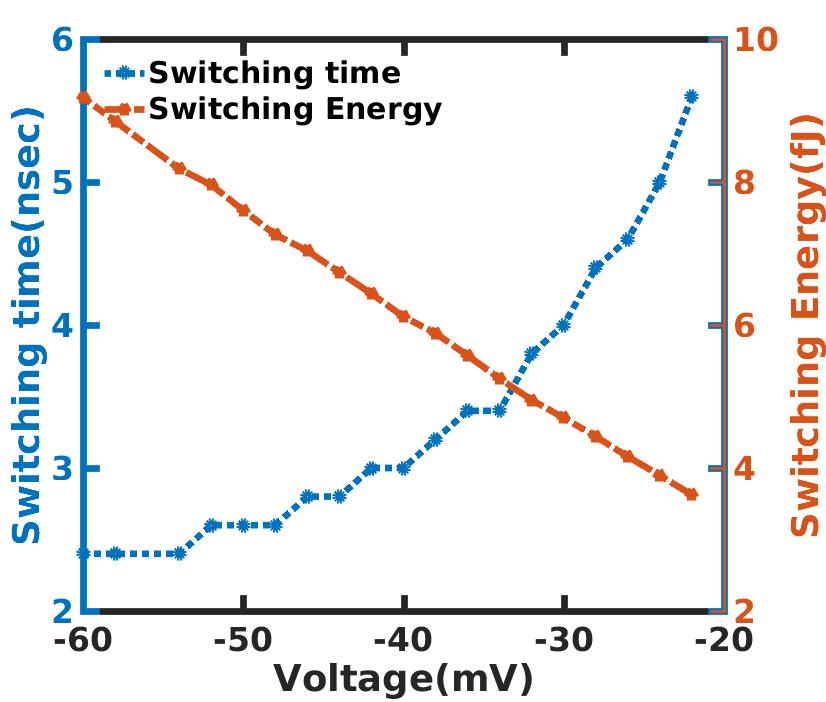}		
	}\subfigure[]{\includegraphics[width=1.75in]{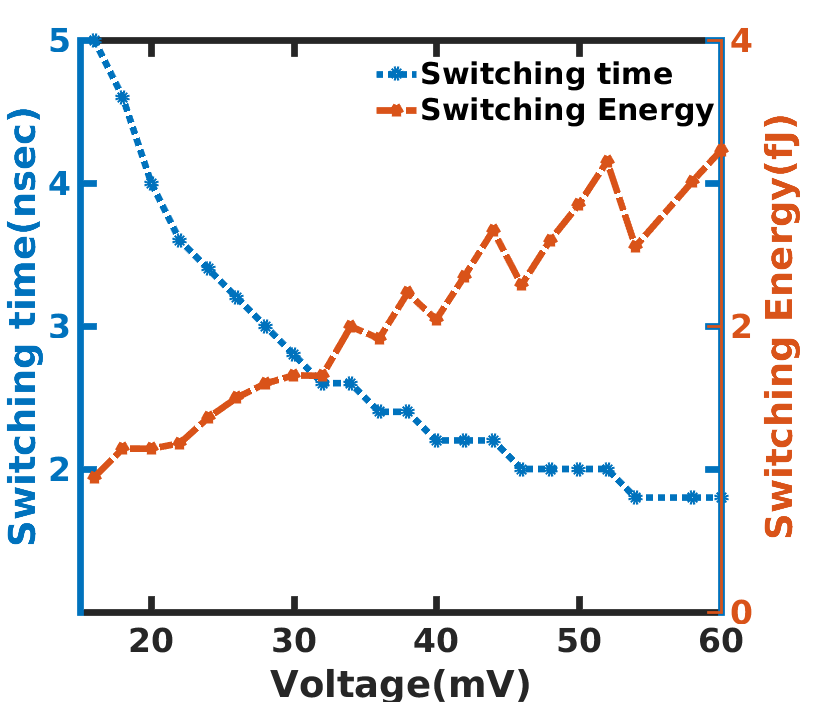}}
	\caption{Switching time and switching energy of p-BPMTJ based MRAM cell from (a) the PC to APC and (b) the APC to PC of the fixed and free FMs}
	\label{ARC_SW_P_E_VV}
\end{figure}

We have used stochastic LLGS equation for magnetization dynamics of the free FM. All the devices have been kept at 300K. The thermal energy due to finite temperature makes the switching process near critical switching bias a stochastic process with fractional switching probabilities. We have removed the zero bias exchange field in all the simulations of MRAM which can be achieved experimentally by applying a static magnetic filed. We have taken 5000 iterations to evaluate the switching probability and switching energy. We have applied different bias voltages with along with different pulse widths ($\tau$) as shown in the Fig.~\ref{chap6_device_design}(a) to evaluate the voltage-pulse width diagram for the switching probability and switching energy.\\
\indent We show in the Fig.~\ref{MTJ_SW_P_tau}(a) \& (b) the switching probability with pulse width or pulse duration from PC to APC and from APC to PC, respectively at different applied bias voltages. The switching probability is of a sigmoidal form as switching bias is larger than critical bias required for switching\cite{He2007,Bedau2010}. As the bias increases, the switching probability also increases due to large spin current associated with higher bias voltage. It can be inferred from the Fig.~\ref{ARFPMTJ_SW_P_tau}(a) and (b), that the sigmoidal switching probability curve with pulse duration can be achieved at lower bias voltages in the BPMTJ based MRAM device in comparison to the MTJ based MRAM device due to a large spin current associated in the BPMTJ structure.\\
\indent We show in the Fig.~\ref{MTJ_SW_P_E_VV_tau}(a), voltage-pulse width diagram of the switching probability for an MTJ based MRAM device from the PC to APC configuration. The fractional switching probability signifies the role of thermal energy for magnetization switching. The switching probability increases and reaches around unity for higher bias voltage and for longer duration of the voltage pulse. The high voltage bias and large pulse duration also increases the energy cost of switching  as shown in the Fig.~\ref{MTJ_SW_P_E_VV_tau}(b). It can be inferred from Fig.~\ref{MTJ_SW_P_E_VV_tau}(c) that, in the APC to PC switching, a higher switching probability can be achieved at the same cost of switching energy as shown in the Fig.~\ref{MTJ_SW_P_E_VV_tau}(d) in comparison to the PC to APC switching due to the asymmetric nature of the spin transfer torque in a trilayer MTJ device(Fig.~\ref{MTJ_MRAM_IV}(c)). It must be noted that for the MTJ based MRAM device, the energy consumption and switching latency in our simulation platform are of the same order as has been reported in different works\cite{Cai2017,Yu2016,Endoh2016}.\\
\indent Figure \ref{ARC_SW_P_E_VV_tau}(a) shows the switching probability diagram for BPMTJ based MRAM cell from the PC to APC. The large spin current due to resonant spin filtering\cite{Sharma2018} in the BPMTJ device reduces the switching bias resulting in lower switching energy as compared to the MTJ based MRAM cell, as shown in \ref{ARC_SW_P_E_VV_tau}(b). We show in the Fig.~\ref{ARC_SW_P_E_VV_tau}(c) and (d), voltage-pulse width diagram of the switching probability  and switching energy from the APC to PC in the BPMTJ based MRAM device.\\ 
\indent We define the switching time as the voltage pulse duration at which switching probability is greater than 0.95. We have extracted the switching time and switching energy for the MTJ and BPMTJ based MRAM devices from the voltage-pulse width diagrams. We show in the Fig.~\ref{MTJ_SW_P_E_VV}(a), switching time and energy with voltage bias for the MTJ based MRAM device from the PC to APC switching of the free FM. The switching time reduces with increase in the strength of bias voltage due to a large spin current and at the same time switching energy also increases. The intersection of switching time and switching energy gives the optimal operating point (OOP) for the MRAM device. The OOP for MTJ based MRAM for the PC to APC switching occurs as $V=-0.26V$ with the switching time ($t_s=3.2nsec$) and switching energy ($E_s=64fJ$). It can be inferred from the Fig.~\ref{MTJ_SW_P_E_VV}(b) that the switching from the APC to PC in the MTJ based MRAM device occurs at lower voltage bias compared to the switching from the PC to APC, due to a large spin current in the positive applied voltage bias(Fig.~\ref{MTJ_MRAM_IV}(a)\&(c)). This shifts the OOP in the MTJ based MRAM in the APC to PC switching towards lower voltage $V=0.18V$ at the same time having lower switching time ($t_s=2.6nsec$) and switching energy ($E_s=24fJ$) in comparison to the switching from the PC to APC. The switching time can further be reduced at the cost of switching energy in the MTJ based MRAM device. Figure \ref{ARC_SW_P_E_VV}(a) shows the switching time and switching energy requirement for the BPMTJ based MRAM device for the PC to APC switching. The OOP for PC to APC switching occurs at the voltage $V=-34mV$ with the switching time ($t_s=3.4nsec$) and switching energy ($E_s=5.2fJ$). We show in \ref{ARC_SW_P_E_VV}(b), the switching time and switching energy variation with bias voltage for the APC to PC switching. The OOP voltage $V=32mV$ for the APC to PC switching is close to the PC to APC switching voltage due to the near symmetric nature of spin current in the BPMTJ device (Fig.~\ref{ARC_SLTMR_MRAM_V}(c)). The OOP switching time ($t_s=2.6nsec$) and switching energy ($E_s=1.7fJ$) for the APC to PC switching is lower in comparison to the the PC to APC switching. We can infer that for nearly the same switching time, the BPMTJ based MRAM is approximately 1170\% and 1370\% more energy efficient than the MTJ based MRAM device in the PC to APC and the APC to PC switching, respectively. The energy efficient switching in the BPMTJ based MRAM devices is attributed to the physics of band-pass spin filtering in these devices.
\section{Conclusions}
\indent We have proposed and explored the design of an MRAM device based on band-pass spin filtering to harness the capability to exhibit a large spin current. We have shown that the physics of band-pass spin filtering makes the BPMTJ a suitable candidate for energy efficient MRAM device. We have estimated that MRAM based on BPMTJ are nearly 1100\% energy efficient in comparison to the trilayer MTJ based MRAM. We believe that the viable device designs presented here and a huge thrust for MRAM as next-generation memory technology will open up new frontiers for experimental considerations of BPMTJ structures.   

\section*{Acknowledgements}
The Research and Development work undertaken in the project
under the Visvesvaraya Ph.D. Scheme of Ministry of Electronics and Information Technology, Government of India, being
implemented by Digital India Corporation (formerly Media
Lab Asia). This work was also supported by the Science and
Engineering Research Board (SERB) of the Government of
India under Grant No. EMR/2017/002853.

\bibliographystyle{IEEEtran}
\bibliography{refSTT}
\end{document}